\documentclass[CTAC, short]{cta-v2mod} 

\usepackage{nicefrac}	
\usepackage{amssymb}	
\usepackage{float}		
\usepackage{lineno}		

\usepackage{multirow}	
\usepackage{subcaption}	
\usepackage{textcomp}	
\usepackage{array}		
\usepackage{url}
\usepackage{pdflscape}	
\usepackage[backgroundcolor = red]{todonotes}		
\usepackage{textgreek}	  
\usepackage{color, soul} 
\usepackage{listings}	 

\usepackage[numbers]{natbib}  
\usepackage{aas_macros}
\usepackage{pdfpages}

\usepackage{chngcntr}
\counterwithout{figure}{chapter}

\usepackage{longtable}

\newcolumntype{H}{>{\columncolor{ctablue}}c}
\newcolumntype{S}{>{\columncolor{lightgray}}c}

\begin{document}
\title{Probing extreme environments with the Cherenkov Telescope Array}
\shorttitle{Probing extreme environments with CTA}
\issue{1}
\revision{a}

\thispagestyle{empty}
\includepdf[pages=1,pagecommand={},scale=1, offset=0cm 0cm]{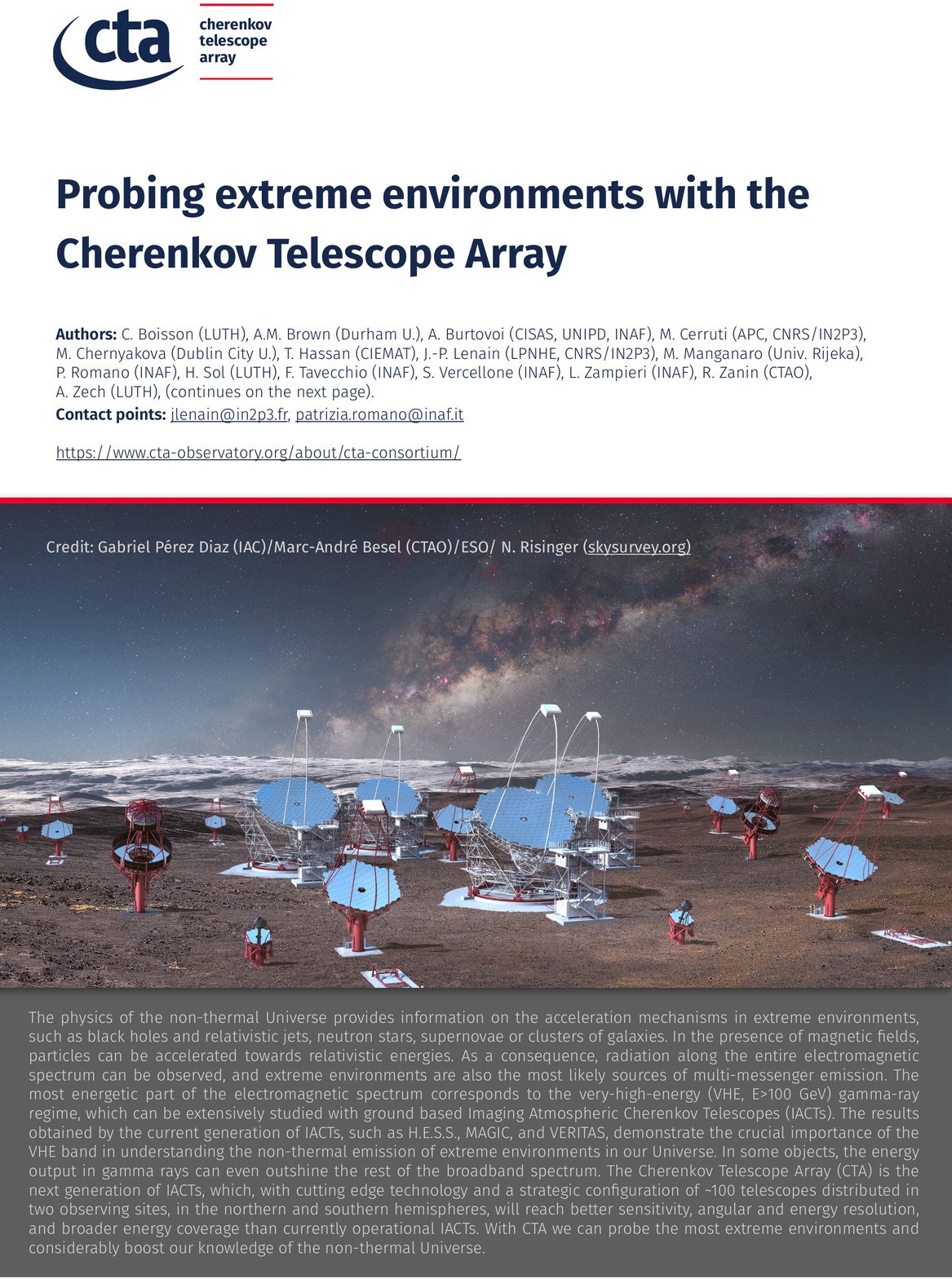}
\clearpage

\FloatBarrier \if@openright\cleardoublepage\else\clearpage\fi

%
{\footnotesize
I. Agudo (Instituto de Astrof\'isica de Andaluc\'ia-CSIC, Spain),
R. Alves Batista (Radboud University Nijmegen, Netherlands),
E.O. Anguner (CPPM, IN2P3/CNRS, France),
L.A. Antonelli (INAF, Osservatorio Astronomico di Roma, Italy),
M. Backes (University of Namibia, Namibia),
C. Balazs (Monash University, Australia),
J. Becerra Gonz\'alez (Instituto de Astrof\'isica de Canarias \& Universidad de La Laguna, Spain),
C. Bigongiari (INAF, Osservatorio Astronomico di Roma, Italy),
E. Bissaldi (Politecnico and INFN Bari, Italy),
J. Bolmont (LPNHE, CNRS/IN2P3, Sorbonne Université, France),
P. Bordas (Universitat de Barcelona, ICCUB, IEEC-UB, Spain),
\v{Z}. Bo\v{s}njak (Zagreb University - FER, Croatia),
M. B\"ottcher (North-West University, South Africa),
M. Burton (Armagh Observatory and Planetarium, UK),
F. Cangemi (LPNHE, CNRS/IN2P3, Sorbonne Université, France),
P. Caraveo (INAF-IASF, Italy),
M. Cardillo (INAF-IAPS, Roma, Italy),
S. Caroff (LAPP, Univ. SMB, CNRS/IN2P3, France),
A. Carosi (University of Geneva, Switzerland),
S. Casanova (Institute of Nuclear Physics Polish Academy of Sciences, Krakow, Poland),
S. Chaty (University of Paris and CEA Paris-Saclay, France),
J.L. Contreras (EMFTEL \& IPARCOS, Universidad Complutense de Madrid, Spain),
J.G. Coelho (UFPR - Universidade Federal do Paran\'a, Brazil),
G. Cotter (University of Oxford, UK),
A. D'A\`i (INAF, IASF Palermo, Italy),
F. D'Ammando (INAF, Istituto di Radioastronomia di Bologna, Italy),
E.M. de Gouveia Dal Pino (IAG-USP, Universidade de S\~ao Paulo, Brazil),
D. de Martino (INAF, Osservatorio Astronomico di Capodimonte-Napoli, Italy),
C. Delgado (CIEMAT, Spain),
D. della Volpe (University of Geneva, Switzerland),
A. Djannati-Ata\"i (Université de Paris, CNRS/IN2P3, APC, France),
R.C. Dos Anjos (UFPR - Universidade Federal do Paran\'a, Brazil),
E. de O\~na Wilhelmi (DESY-Zeuthen, Germany),
V. Dwarkadas (University of Chicago, USA),
G. Emery (DPNC - University of Geneva, Switzerland),
E. Fedorova (Taras Shevchenko National University of Kyiv, Ukraine),
S. Fegan (LLR/Ecole Polytechnique, CNRS/IN2P3, France),
M.D. Filipovic (Western Sydney University, Australia),
G. Galanti (INAF - IASF Milano, Italy),
D. Gasparrini    (INFN Roma Tor Vergata \& ASI-SSDC, Italy),
G. Ghirlanda (INAF, Osservatorio Astronomico di Brera, Italy),
P. Goldoni (APC/IRFU, France),
J. Granot (Open University of Israel, Israel),
J.G. Green (INAF, Osservatorio Astronomico di Roma, Italy),
M. Heller (University of Geneva, Switzerland),
B. Hnatyk (Taras Shevchenko National University of Kyiv, Ukraine),
R. Hnatyk (Taras Shevchenko National University of Kyiv, Ukraine),
D. Horan (LLR/Ecole Polytechnique, CNRS/IN2P3, France),
T. Hovatta (FINCA, University of Turku, Finland),
S. Inoue (RIKEN, Japan),
M. Jamrozy (Jagiellonian University, Poland),
B. Kh\'elifi (APC, IN2P3/CNRS - Université de Paris, France),
N. Komin (University of the Witwatersrand, Johannesburg, South Africa),
K. Kohri (KEK, Japan),
A. Lamastra (INAF, Osservatorio Astronomico di Roma, Italy),
N. La Palombara (INAF, IASF Milano, Italy),
E. Lindfors (FINCA, University of Turku, Finland),
I. Liodakis (FINCA, University of Turku, Finland),
S. Lombardi (INAF-Osservatorio Astronomico di Roma, Italy),
F. Longo (University and INFN Trieste, Italy),
F. Lucarelli (INAF-Osservatorio Astronomico di Roma \& ASI-SSDC, Italy),
P.L. Luque-Escamilla (University of Jaen, Spain),
J. Marti (University of Jaen, Spain),
M. Martinez (Institut de F\'isica d'Altes Energies, IFAE-BIST, Barcelona, Spain),
D. Mazin (ICRR, University of Tokyo, Japan and MPI for Physics, Munich, Germany),
J.D. Mbarubucyeye (DESY-Zeuthen, Germany),
S. Menchiari (Universit\`a degli studi di Siena, Italy),
L. Mohrmann (FAU Erlangen-N\"urnberg, Germany),
T. Montaruli (University of Geneva, Switzerland),
A. Moralejo Olaizola (Institut de F\'isica d'Altes Energies, IFAE-BIST, Barcelona, Spain),
D. Morcuende (EMFTEL \& IPARCOS, Universidad Complutense de Madrid, Spain),
G. Morlino (INAF-Osservatorio Astrofisico di Arcetri, Italy),
A. Morselli (INFN Roma Tor Vergata),
C.G. Mundell (University of Bath, UK),
T. Murach (DESY Zeuthen, Germany),
A. Nagai (University of Geneva, Switzerland),
A. Nayerhoda (Institute of Nuclear Physics Polish Academy of Sciences, Krakow, Poland),
J. Niemiec (Institute of Nuclear Physics Polish Academy of Sciences, Krakow, Poland),
M. Nikolajuk (University of Bialystok, Poland),
B. Olmi (INAF, Osservatorio Astronomico di Palermo, Italy),
M. Orienti (INAF, Istituto di Radioastronomia di Bologna, Italy),
J.M. Paredes (Universitat de Barcelona, ICCUB, IEEC-UB, Spain),
G. Pareschi (INAF, Osservatorio Astronomico di Brera, Italy),
A. Pe'er (Bar Ilan University, Israel, and University College Cork, Ireland),
G. P\"uhlhofer (IAAT, University of T\"ubingen, Germany),
M. Punch (APC, IN2P3/CNRS - Université de Paris, France),
O. Reimer (Innsbruck University, Institute for Astro- and Particle Physics, Austria),
M. Rib\'o (Universitat de Barcelona, ICCUB, IEEC-UB, Spain),
F. Rieger (Max-Planck-Institut f\"ur Kernphysik, Heidelberg, Germany),
G. Rodriguez (INAF-IAPS, Roma, Italy),
J. Rodriguez (AIM, CEA, CNRS, Université Paris-Saclay, Université de Paris, France),
G. Romeo (INAF, Osservatorio Astrofisico di Catania, Italy),
M. Roncadelli (INFN - Pavia, Italy),
G. Rowell (University of Adelaide, Australia),
B. Rudak (CAMK PAN, Poland),
I. Sadeh (DESY-Zeuthen, Germany),
F. Salesa Greus (Institute of Nuclear Physics Polish Academy of Sciences, Krakow, Poland),
U. Sawangwit (National Astronomical Research Institute of Thailand, Thailand),
F. Sch\"ussler (IRFU, CEA Paris-Saclay),
O. Sergijenko (Taras Shevchenko National University of Kyiv, Ukraine),
R.C. Shellard (CBPF, Rio de Janeiro, Brazil),
F.G. Saturni (INAF, Osservatorio Astronomico di Roma, Italy),
A. Stamerra (INAF, Osservatorio Astronomico di Roma, Italy),
Th. Stolarczyk (AIM, CEA, CNRS, Université Paris-Saclay, Université de Paris, France),
G. Tagliaferri (INAF, Osservatorio Astronomico di Brera, Italy),
V. Testa (INAF, Osservatorio Astronomico di Roma, Italy),
L. Tibaldo (IRAP, Université de Toulouse, CNRS, UPS, CNES, Toulouse, France),
S. Ventura (University of Siena - INFN Pisa, Italy),
A. Viana (IFSC-USP, Universidade de S\~ao Paulo, Brazil),
J. Vink (Anton Pannekoek Institute/GRAPPA, University of Amsterdam, Netherlands),
V. Vitale (INFN, Roma Tor Vergata, Italy),
S. Vorobiov (University of Nova Gorica, Slovenia),
A. Wierzcholska (Institute of Nuclear Physics Polish Academy of Sciences, Krakow, Poland),
M. Zacharias (LUTH, Observatoire de Paris, France),
D. Zavrtanik (University of Nova Gorica, Slovenia),
for the CTA consortium.
}

\chapter*{Introduction} 

Most cosmic objects such as stars and dust, produce thermal radiation that dominates the IR and optical Universe. However, several exceptional sources release non-thermal radiation, often extending over the entire electromagnetic spectrum, from the radio to the VHE gamma-ray band. The non-thermal Universe is not limited to photons, but it also includes other particles, such as cosmic rays or neutrinos, which can only be produced in the most violent and extreme environments. The most intriguing non-thermal sources are perhaps those able to emit photons with energies in the gamma-ray band, flagging the existence of particles accelerated up to ultra-relativistic energies.

Many experiments and observatories in the past years focused on the study of the most energetic part of the electromagnetic spectrum, to investigate the extreme processes that can accelerate particles to such energies and produce high-energy and VHE gamma rays, neutrinos and cosmic rays. Nowadays more than 200 sources\footnote{\url{http://tevcat.uchicago.edu/}, \cite{2008ICRC....3.1341W}} have been found to emit VHE gamma rays and almost thirty times more to emit high-energy (100\,MeV--100\,GeV) gamma rays~\cite{2020ApJ...892..105A}. Gamma-ray data, coupled with multi-frequency observations, provide a unique tool to probe the physical processes acting in these extreme sources. The emission is often variable, sometimes showing drastic luminosity changes on very short time scales (down to minutes), making the explanation of the physical mechanisms even more challenging.

This white paper is dedicated to the physics of the non-thermal Universe with VHE gamma rays. We aim to demonstrate how CTA will be able to probe the most extreme environments of our Universe thus addressing key issues such as the identification of the locations of the gamma-ray emitting zones in active galactic nuclei (AGNs), determining the nature of the radiating particles and the acceleration mechanisms at work, and understanding the origins of the variability observed in most energy bands, thus improving our knowledge of AGN populations and their unification schemes. A description of the cutting edge technology adopted by CTA and the expected performance of the two arrays can be found in \cite{2019scta.book.....C}. 

\chapter{Extreme astrophysical environments} 
\label{env}

The production of energetic photons and particles is often associated with the presence of compact objects, such as neutron stars or black holes, and powerful outflows or jets. VHE gamma-ray emission from accelerated particles can therefore act as a probe of these environments, in different ways. First, the study of the spectrum of AGNs in a broadband context can allow us to identify the mechanism of the emission, while variability helps measuring the size of the emission region. The investigation of correlated flux variability observed at different wavelengths can further shed light on the evolution of the parent particle energy distribution.

In addition, the extreme energy that the gamma rays can achieve allows the production of electron-positron pairs, a process that can happen only in the high-energy domain. The propagation of VHE gamma rays across the Universe can reveal key information on the structure of the Universe itself. For instance, VHE gamma rays can be used to probe magnetic fields in cosmic voids down to values many orders of magnitude below the reach of any other technique \cite{2021JCAP...02..048A}. Gamma-ray observations can also establish if VHE photons heat the gas in these under-dense regions, suppressing the formation of dwarf satellite galaxies.

\chapter{Black holes and jets} 
\label{BH}

About 15\% of the population of AGNs discovered so far have relativistic jets \cite{2017A&ARv..25....2P}. When pointing at us, the jet emission is strongly boosted due to relativistic beaming. These sources, called blazars, represent the majority of the objects detected in the gamma-ray range \cite{2017MNRAS.469..255G}.

\begin{figure}
\centering
    \includegraphics[scale=0.8]{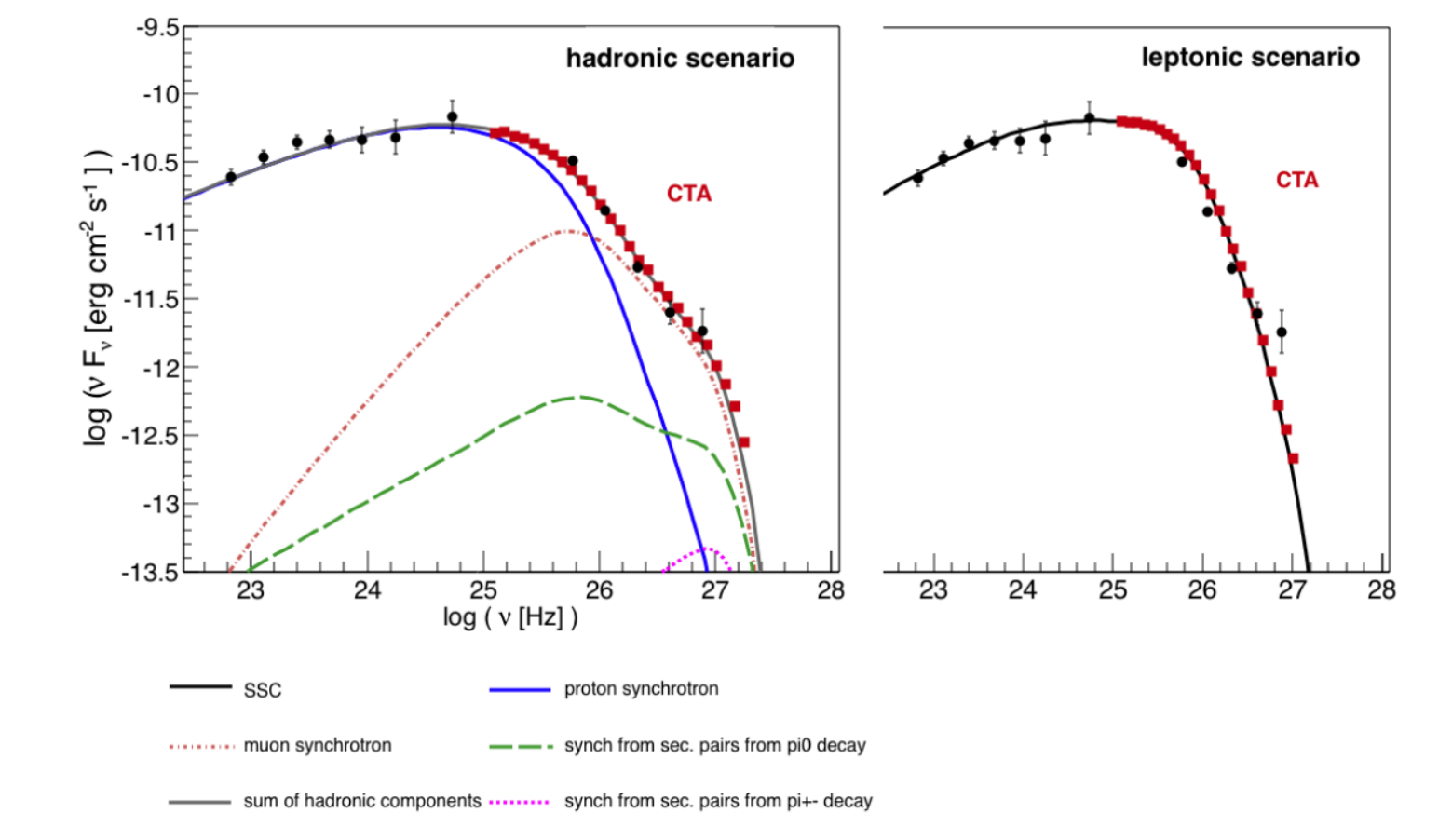}
    \caption{A comparison of the expected CTA spectra for two specific (simple) emission models for the blazar PKS~2155$-$304. A hadronic scenario, where high-energy emission is caused by proton- and muon-synchrotron photons and secondary emission from proton-photon interactions, is shown on the left, and a standard leptonic synchrotron self-Compton model on the right. The exposure time assumed for the simulations (33\,h) is the same as the live time for the H.E.S.S.\ observations (black data points above 3$\times10^{25}$\,Hz). 
    From~\cite{2019scta.book.....C}, adapted from \cite{2017A&A...602A..25Z}.
    }
    \label{fig:models}
\end{figure}

The extreme environment displayed in black hole-jet systems provides appropriate conditions for particle acceleration to the highest energies. Efficient particle acceleration occurs in jets from the innermost portions up to the terminal regions (i.e.\ Mpc scale, in the most powerful AGNs). The mechanisms accelerating particles (leptons, possibly nuclei) to ultra-relativistic energies are still a matter of intense investigation  \cite{2019ARA&A..57..467B,2020LRCA....6....1M}. 
Similarly, different mechanisms are active along the jets \cite{2020NewAR..8901543M}, with magnetic reconnection (possibly induced by fluid instabilities) prevailing at the smallest scales (where, according to the current theoretical framework, see e.g.\ \cite{2007MNRAS.380...51K}, the jet energy flux is largely dominated by  magnetic fields), and diffusive shock acceleration mostly active at large, matter dominated scales, where internal shocks can form due to disturbances in the jet or oblique shocks can be induced by reconfinement by external gas. Recent numerical efforts (e.g. \cite{2014ApJ...783L..21S, 2016ApJ...816L...8W}) show that relativistic magnetic reconnection can produce non-thermal particle distributions that fit the observed emission and account for the surprisingly rapid (down to few minutes) VHE flares occasionally recorded in blazars \cite{2019ApJ...880...37P}. CTA will distinguish between the different possible acceleration scenarios. A key question concerns the potential role of jets and black holes in the production of ultra-high-energy cosmic rays. The potential association of a blazar with a high-energy neutrino \cite{2018Sci...361.1378I} hints at the existence of nuclei with energies of 100\,PeV. 

To effectively test our theoretical framework, it is crucial to model the emission from such environments in a broadband context using data collected over the entire electromagnetic spectrum. A fundamental role is played by observations in the VHE band, a unique window to access the emission from the most energetic particles. Even if the high-energy bump of the spectral energy distribution (SED) could be explained by Compton scattering of the same population of electrons that radiates in the synchrotron bump (simple leptonic scenario), in reality broadband studies often require the introduction of other radiative components, hadronic or hybrid (lepto-hadronic) \cite{2019MNRAS.483L..12C}, or of multiple emission zones \cite{2019ApJ...886...23X,2020A&A...640A.132M} to explain the broadband experimental results. Observations at VHE with CTA will be essential in the effort of testing and constraining the models, making it possible, for instance, to trace the tiny spectral differences expected in the different radiative scenarios (Fig.~\ref{fig:models}).

The lower energy threshold of CTA will enable us not only to detect in VHE more distant sources (the most distant AGN detected in this energy range is located at a redshift $z$=0.954 \cite{2016A&A...595A..98A}) but also extra-galactic objects belonging to relatively new classes of jetted sources, such as narrow-line Seyfert galaxies (NLSy1). The detection of VHE emission from NLSy1s  \cite{2018MNRAS.481.5046R,2020MNRAS.494..411R} could provide crucial insights into the unification of various classes of AGNs and on the conditions for jet formation as a function of the supermassive black hole mass and accretion rate, as well as the role of the host galaxy. 

Furthermore, an extended VHE signal has recently been detected from the radio-galaxy Cen\,A \cite{2020Natur.582..356H}. CTA will allow us to provide more stringent information on the location of the VHE emission region(s) and acceleration mechanism in relativistic jets, in complementarity with highly resolved images of jets from radio VLBI observations, from Cen\,A, M\,87, and other nearby radio galaxies.

\begin{figure}
   \centering
   \includegraphics[width=0.38\textwidth]{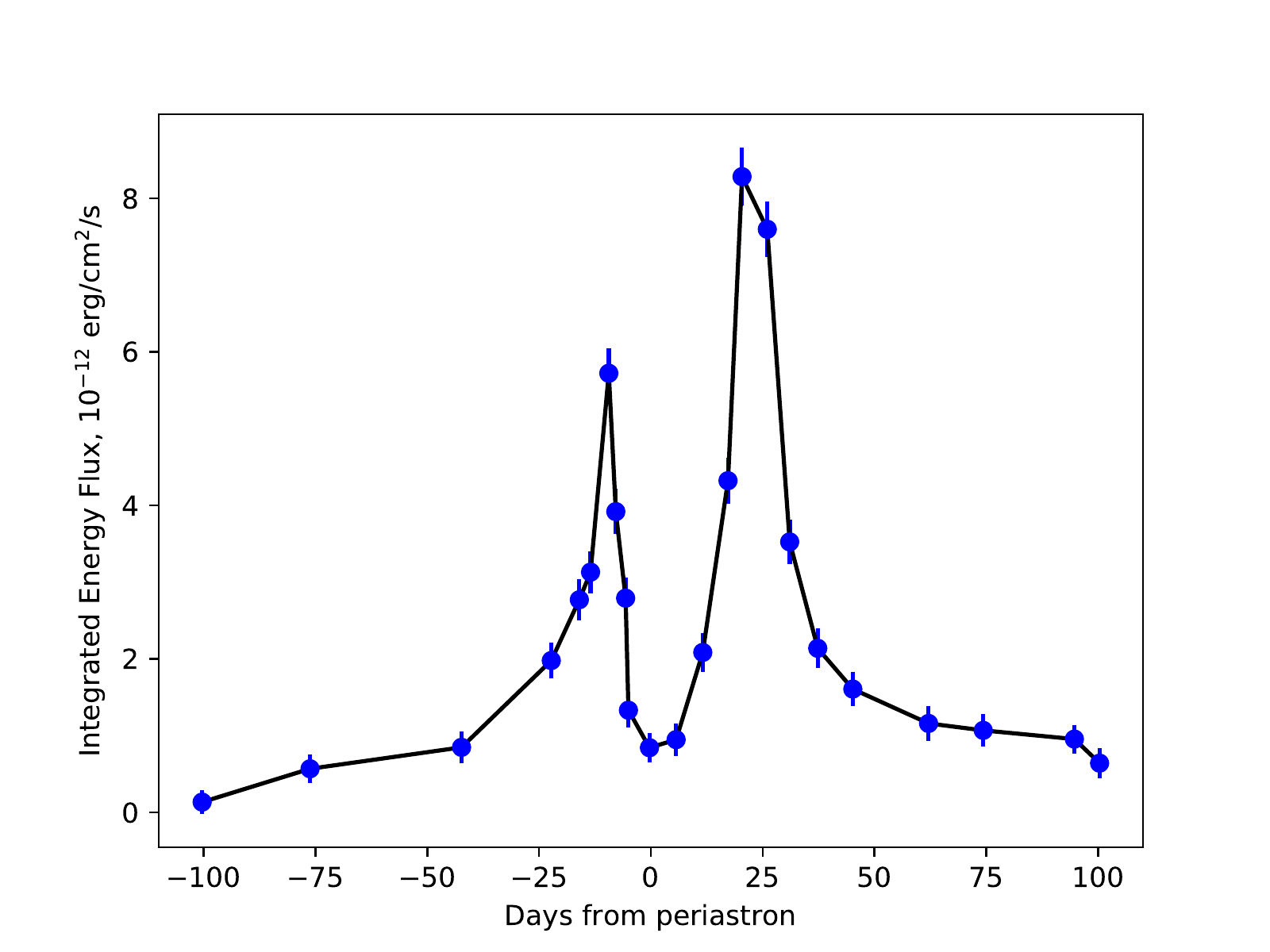}
   \includegraphics[width=0.60\textwidth]{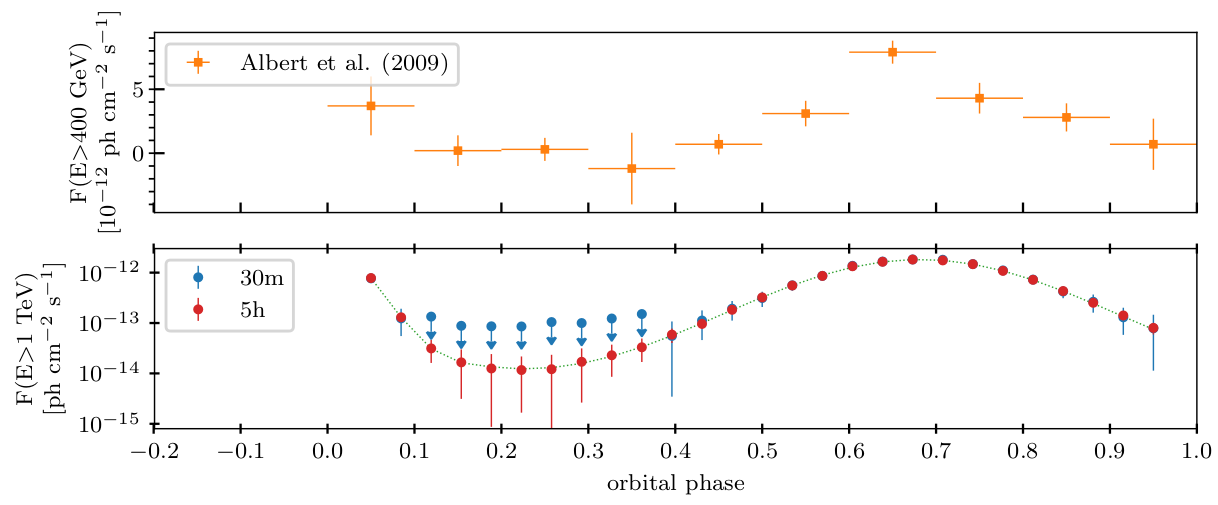}
   \caption{\textit{Left panel:} Simulated  light  curve  above 1\,TeV  of  PSR~B1259$-$63 around the periastron. Each point has a 30\,min exposure. 
   \textit{Right panel:} Comparison of the MAGIC observations of LSI $+$61$^\circ$ 303 \cite{Albert2009} with the simulations of CTA observations with 30\,min and 5\,h exposure. From  \cite{2019A&A...631A.177C}.
   } 
    \label{fig:gbin}
\end{figure}

\chapter{Neutron stars and relativistic outflows} 
\label{GAL}

So far, VHE pulsed emission has been detected with IACTs from four pulsars \citep{2011Sci...334...69V, 2018A&A...620A..66H, 2019arXiv190806464S, 2020A&A...643L..14M}, but only two of them – the Crab and Vela pulsars – show significant emission at TeV energies. Observing these sources with CTA will make it possible to better characterize their VHE emission and to probe the physical conditions in the surroundings of neutron stars, allowing us to determine the maximum energy of the accelerated particles, to constrain the geometry and location of the emission region, and to investigate the physical mechanism(s) that produce pulsations at TeV energies.

CTA will also play a key role in studying the details of colliding winds in gamma-ray binaries and micro-quasars, binary systems visible up to TeV energies, with about a hundred such objects expected to be detected \cite{2017A&A...608A..59D}, while less than 10 such systems are currently known \cite{Chernyakova2020review}. It is indeed of paramount importance to understand what makes particle acceleration in these systems so effective and to discover more objects of such a kind. The  unprecedented  sensitivity of CTA will allow us to measure the spectral parameters of these systems on time scales comparable to the variability time observed at other energies, as illustrated in Fig.~\ref{fig:gbin} using as an example PSR~B1259$-$63 (left panel) and LSI $+$61$^\circ$ 303 (right panel). This will allow us to discriminate between the existing models, and provide input data for theories describing details of particle acceleration \cite{2019A&A...631A.177C}.

\chapter{Gamma-ray propagation} 
\label{icons/propa}

Owing to its unique capabilities, observations with CTA will drastically improve the studies on the VHE gamma-ray propagation, enabling measurements of the extragalactic background light (EBL) and of the cosmic gamma-ray horizon (CGRH), the limit beyond which the Universe is opaque to VHE gamma rays, with unprecedented precision. Gamma rays interact via electron-positron pair production with photons from IR to UV which fill our Universe and constitute the EBL. The flux of VHE gamma rays that we can detect on Earth is then suppressed with respect to the intrinsic emission \cite{2012MNRAS.422.3189G,2008A&A...487..837F,2011MNRAS.410.2556D}. The detection of distant sources is limited by this process: CTA, with its lower energy threshold and increased sensitivity as compared with operating IACTs, will be able to greatly enlarge the horizon.

The EBL has been measured using gamma-ray data from stacked samples on several  \cite{2012Sci...338.1190A,2013A&A...550A...4H} or individually from the furthest detected blazars \cite{2019MNRAS.486.4233A} and the transparency of the Universe has been tested. With even more distant sources, those measurements can be further improved, and it will also be possible to obtain a new measurement of the CGRH \cite{2013ApJ...770...77D} and of the expansion rate of the Universe \cite{2013ApJ...771L..34D} (see also \cite{2021JCAP...02..048A} for further details).

Moreover, the investigation of gamma-ray propagation can probe extremely weak magnetic fields thought to exist in cosmic voids \cite{2021JCAP...02..048A}. In fact, the pairs produced by the interaction with the EBL can further interact with the Cosmic Microwave Background (CMB) photons, and generate a secondary gamma-ray flux which could be detected around blazars and other types of AGNs. This process can be observed in two ways, one being the presence of a low-energy spectral component on top of the observed point-sources spectra, and the other one being an extended gamma-ray halo around distant blazars. The properties of such extended emission depend on the intergalactic magnetic field strength and coherence length: CTA will have the possibility to measure such effects and simulations have shown that 50 hours of data would be sufficient to detect the putative halo around the blazar 1ES~0229$+$200 \cite{2021JCAP...02..048A}.

\chapter*{Conclusions} 
\label{conclusions}

The future of high-energy astrophysics lies in the multi-messenger context, in which information collected with photons across the entire electromagnetic spectrum, neutrinos, cosmic rays and gravitational waves can be combined in order to increase our knowledge of the non-thermal Universe. Gamma rays are an important part of this picture, and observations in the VHE gamma-ray range are necessary more than ever. In the past years, many impressive results from IACTs have clearly demonstrated how the multi-wavelength study of astrophysical objects is a fundamental tool for the study of our Universe. CTA will allow us to study extreme astrophysical environments with unprecedented sensitivity and resolution in the most energetic part of the electromagnetic spectrum, producing more exciting results to be shared with the astrophysics community. More information on the CTA project can be found in \cite[and references therein]{2019scta.book.....C,2013APh....43....1H} and on the CTA web page (\href{https://www.cta-observatory.org/}{https://www.cta-observatory.org/}). 

\textbf{See also the other white papers sent to the ASTRONET committee on behalf of the CTA consortium.}

{\em Acknowledgements.}
We gratefully acknowledge financial support from the agencies and organizations listed here: \href{http://www.cta-observatory.org/consortium_acknowledgments}{http://www.cta-observatory.org/consortium\_acknowledgments}.

\bibliographystyle{apsrev}
\bibliography{main.bbl}

\end{document}